\documentclass[11pt]{article}

\usepackage{caption}
\usepackage{subcaption}

\usepackage{comment}
\usepackage{amssymb}
\usepackage{amsmath}
\usepackage{amsfonts}
\usepackage{mathrsfs}
\usepackage{engord}
\usepackage[dvips]{graphicx}
\usepackage{bbm}
\usepackage{threeparttable}
\usepackage{booktabs}
\usepackage{epsfig}
\usepackage{color}
\usepackage{graphicx}
\usepackage{multirow}
\usepackage{cite}
\usepackage{epstopdf}
\usepackage{amsthm}
\usepackage{appendix}
\usepackage{framed}
\usepackage{url}

\usepackage{color}

\begin{document}

\newtheorem{thm}{Theorem}
\newtheorem{prop}{Proposition}
\newtheorem{lem}{Lemma}
\newtheorem{defn}{Definition}
\newtheorem{ex}{Example}
\newtheorem{cor}{Corollary}
\newtheorem{prn}{Principle}
\newtheorem{case}{Case}
\newtheorem{rmk}{Remark}
%
\title{Synthesis of Gaussian Trees with Correlation Sign Ambiguity: An Information Theoretic Approach}

\author{Ali Moharrer,
Shuangqing Wei,
George T. Amariucai, 
and Jing Deng}
\maketitle
\footnotetext[1]{A. Moharrer, and S. Wei are with the school of Electrical Engineering and Computer Science, Louisiana State University, Baton Rouge, LA 70803, USA (Email: amohar2@lsu.edu, swei@lsu.edu). 

G. T. Amariucai is with the department of Electrical and Computer Engineering, Iowa State University, Ames, IA, USA (Email: gamari@iastate.edu). 

J. Deng is with the department of Computer Science, University of North Carolina at Greensboro, Greensboro, NC, USA (Email: jing.deng@uncg.edu).

This material is based upon work supported in part by the National Science
Foundation under Grant No. 1320351.}



\begin{abstract}
In latent Gaussian trees the pairwise correlation signs between the variables are intrinsically unrecoverable.
Such information is vital since it completely determines the direction in which two variables are associated. 
In this work, we resort to information theoretical approaches to achieve two fundamental goals: 
First, we quantify the amount of information loss due to unrecoverable sign information. 
Second, we show the importance of such information in determining the maximum achievable rate region, in which the observed output vector can be synthesized, given its probability density function.
In particular, we model the graphical model as a communication channel and propose a new layered encoding framework to synthesize observed data using upper layer Gaussian inputs and independent Bernoulli correlation sign inputs from each layer. 
We find the achievable rate region for the rate tuples of multi-layer latent Gaussian messages to synthesize the desired observables.
\end{abstract}

\section{Introduction}

Let $\mathbf{X}=\{X_1,X_2,...,X_n\}$ be the $n$ observed variables, while the set of variables $\mathbf{Y}=\{Y_1,Y_2,...,Y_k\}$ are hidden to us. The goal of any inference algorithm is to recover the hidden parameters related to those $k$ hidden nodes ($k$ may be unknown).
Consider a special subset of graphical models, known as latent \textit{Gaussian trees}, in which the underlying structure is a tree and the joint density of the variables is captured by a Gaussian density.
The Gaussian graphical models are widely studied in the literature because of a direct correspondence between conditional independence relations occurring in the model with zeros in the inverse of covariance matrix, known as the \textit{concentration matrix}.

There are several works such as \cite{mit,nj} that have proposed efficient algorithms to infer the latent Gaussian tree parameters. In fact, Choi et al., proposed a new \textit{recursive grouping} (RG) algorithm along with its improved version, i.e., \textit{Chow-Liu} RG (CLRG) algorithm to recover a latent Gaussian tree that is both \textit{structural} and \textit{risk} consistent \cite{mit}, hence it recovers the \textit{correct} value for the latent parameters.
They introduced a \textit{tree metric} as the negative \textit{log} of the absolute value of pairwise correlations to perform the algorithm.
Also, Shiers et al., in \cite{correlation}, characterized the correlation space of latent Gaussian trees and showed the necessary and sufficient conditions under which the correlation space represents a particular latent Gaussian tree.
Note that the RG algorithm can be directly related to correlation space of latent Gaussian trees in a sense that it recursively checks certain constraints on correlations to converge to a latent tree with true parameters.

These methods have been shown successful in estimating latent parameters in terms of both computational complexity and consistency.
However, regardless of which inference algorithm is used, complete inference of the correct correlation signs (sign of edge-weights) from the observed data is impossible in latent Gaussian trees, since there is an intrinsic singularity issue with such models. 
It turns out that the sign singularity in latent Gaussian trees is due to the fact that there models in general are \textit{non-identifiable}, whereas, a regular (non-singular) model is defined to be both \textit{identifiable} and having a \textit{positive definite metric} \cite[p. 10]{watanabe}.
\begin{figure} [h!]
\centering 
\includegraphics[scale=0.6]{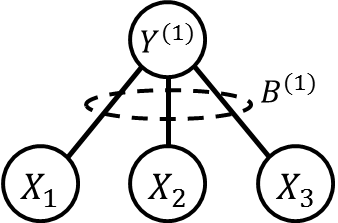}
\caption{A simple Gaussian tree with a hidden node $Y^{(1)}$\label{fig:broadcast}} 
\end{figure}

In this paper, we first resort to information theoretic tools to quantify such information loss in inferred correlation sign values.
As our second step, by modeling the latent Gaussian tree as a multi-layer communication channel, we propose an encoding scheme to generate the observable vector as an output, using the latent and sign variables as the inputs to the channel.
Such layered approach is efficient in a sense that by relying on a latent Gaussian tree structure, it uses smaller number of parameters to generate the output vector. We will discuss such efficiency in more detail in later sections.
Lastly, using the quantified sign information we characterize the achievable rate region of rate tuples for latent sign and latent variables inputs.
To demonstrate our approach, consider a Gaussian tree shown in Figure \ref{fig:broadcast}.
We may think of this Gaussian tree as a communication channel, where information flows from a Gaussian source $Y^{(1)}\sim N(0,1)$ through three communication channels $p_{X_i|Y^{(1)}}(x_i|y^{(1)})$ with independent additive Gaussian noise variables $Z_i\sim N(0,\sigma^2_{z_i}),~i\in\{1,2,3\}$ to generate (dependent) outputs with $\mathbf{X}\sim N(0,\Sigma_{\mathbf{x}})$.
We introduce $B^{(1)}\in\{-1,1\}$ as a binary Bernoulli random variable as another input to the channel, which reflects the sign information of pairwise correlations. 
Define $\rho_{x_iy}=E[X_iY]$ as true correlation values between the input and each of the three output.
For the channel in Figure \ref{fig:broadcast}, one may assume that $B^{(1)}=1$ to show the case with $\rho'_{x_iy}=\rho_{x_iy}$, while $B^{(1)}=-1$ captures $\rho''_{x_iy}=-\rho_{x_iy}$, where $\rho'_{x_iy}$ and $\rho''_{x_iy},~i\in\{1,2,3\}$, are the recovered correlation values using certain inference algorithm such as RG \cite{mit}.
It is easy to see that both recovered correlation values induce the same covariance matrix $\Sigma_{\mathbf{x}}$, showing the sign singularity issue in such a latent Gaussian tree.
Cuff \cite{cuff} introduced a memoryless channel synthesis model through which the output is generated by certain encoding/decoding on channel inputs.
Similarly, in this paper we focus on obtaining achievable rates through which the Gaussian output can be synthesized.
More specifically, our goal is to characterize the achievable rate region and through an encoding scheme to synthesize Gaussian outputs with density $q_\mathbf{X}(\mathbf{x})$ using only Gaussian inputs and through a channel with additive Gaussian noises, where the synthesized joint density $q_\mathbf{X}(\mathbf{x})$ is indistinguishable from the true output density $p_\mathbf{X}(\mathbf{x})$ as measured by \textit{total variation} metric \cite{cuff}.
In particular, we find a solution for  $\inf_{\tilde{\mathbf{Y}}} I(\mathbf{X};\tilde{\mathbf{Y}})$, where $I(\mathbf{X};\tilde{\mathbf{Y}})$ is the mutual information between the output $\mathbf{X}$ and the input vector $\tilde{\mathbf{Y}}=\{\mathbf{Y},\mathbf{B}\}$.
This corresponds to finding the minimum achievable rate to synthesize the Gaussian output.
We show that such quantity is only a function of output joint density. Hence, given output it cannot be further optimized.
However, we show that to obtain the maximum rate region to synthesize the output, one may minimize $I(\mathbf{X};\mathbf{Y})$, which in turn will be equivalent to maximizing the conditional mutual information $I(\mathbf{X};\mathbf{B}|\mathbf{Y})$, hence, showing the maximum amount of lost sign information.
In such settings, we show that the input $\mathbf{B}$ and the output $\mathbf{X}$ are independent, by which we provide another reason on why previous learning approaches \cite{mit,correlation} are incapable of inferring the sign information.

The rest of the paper is organized as follows. Section \ref{sec:formulation}
gives the problem formulation and models the sign singularity problem.
Main results of the paper regarding achievable rate region are discussed in Section \ref{sec:main}.
We conclude the paper in Section \ref{sec:conclusion}.

\section{Problem Formulation} \label{sec:formulation}

\subsection{The signal model of a latent Gaussian tree}
Here, we suppose a latent graphical model, with $\mathbf{Y}=[Y_1,Y_2,...,Y_k]'$ as the set of hidden variables, and $\mathbf{X}=[X_1,X_2,...,X_n]'$ as the set of observed variables.
We also assume that the underlying network structure is a \textit{minimal} latent Gaussian tree \cite{mit}, therefore, making the joint probability $P(X,Y)$ be a Gaussian joint density $N(\mathbf{\mu},\Sigma_{\mathbf{xy}})$, where the covariance matrix $\Sigma_{\mathbf{xy}}$ induces tree structure $G_T(V,E,W)$, where $V$ is the set of nodes consisting of both vectors $\mathbf{X}$ and $\mathbf{Y}$; $E$ is the set of edges; and $W$ is the set of edge-weights determining the pairwise covariances between any adjacent nodes.
In a minimal Gaussian tree we assume all the hidden variables have at least three neighbors \cite{mit}, which results in ignoring all those singular cases where there can be arbitrarily redundant hidden variables added to the model without changing the observed joint density $p_{\mathbf{X}}(\mathbf{x})$.
We consider normalized variances for all variables $X_i\in \mathbf{X},~i\in\{1,2,...,n\}$ and $Y_j\in\mathbf{Y},~j\in\{1,2,...,k\}$. Such constraints do not affect the channel structure, and hence the independence relations captured by $\Sigma_{\mathbf{xy}}$.
Without loss of generality, we also assume $\mathbf{\mu}=\mathbf{0}$, this constraint does not change the amount of information carried by the observed vector. 

\begin{figure} [h!]
\centering 
\includegraphics[scale=0.5]{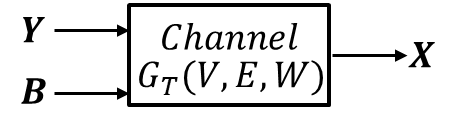}
\caption{A communication system with both sign and latent variables vectors as an input\label{fig:channel2}} 
\end{figure}

In order to quantify the amount of lost sign information we model our problem as shown in Figure \ref{fig:channel2}. 
In fact, we introduce a vector $\mathbf{B}=\{B_1,...,B_m\}$, with each $B_i\in\{-1,1\}$ being a binary Bernoulli random variable with parameter $\pi_i=p(B_i=1)$ as another input to the channel, which captures the sign information of pairwise correlations.
Assume $\mathbf{Y}$ and $\mathbf{B}$ as the input vectors, $\mathbf{X}$ as the output vector, and the noisy channel to be characterized by the conditional probability distribution $P_{\mathbf{X}|\mathbf{Y},\mathbf{B}}(\mathbf{x}|\mathbf{y},\mathbf{b})$, the signal model for such a channel can be written as follows,
\begin{align} \label{eq:linear_regression}
\mathbf{X}=\mathbf{A_BY}+\mathbf{Z}
\end{align}
\noindent where $\mathbf{A_B}$ is $n\times k$ channel gain matrix that also carries the sign information vector $\mathbf{B}$, and $\mathbf{Z}=[Z_1,...,Z_n]'\sim N(0,\Sigma_{\mathbf{z}})$ is the additive noise vector, with a diagonal covariance matrix $\Sigma_{\mathbf{z}}$, where the diagonal entries $\sigma^2_{z_i}$ are the variances of $Z_i$.

\subsection{Studying the properties of sign information vector $\mathbf{B}$}

\begin{figure} [h!]
\centering 
\includegraphics[scale=0.6]{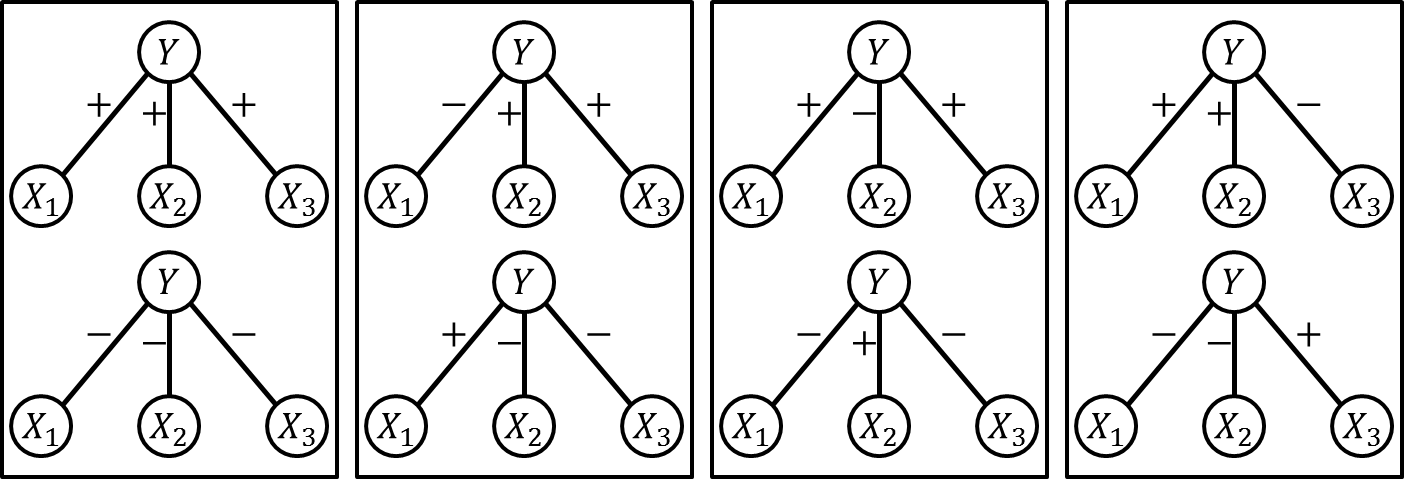}
\caption{Sign singularity in the star Gaussian tree at each group\label{fig:sign}} 
\end{figure}

As an example, consider the channel shown in Figure \ref{fig:broadcast}.
Given enough samples from each of the outputs $X_1$, $X_2$, and $X_3$, one can estimate the pairwise correlations $\rho_{x_ix_j},~i,j\in\{1,2,3\}$ and use existing learning algorithms such as RG \cite{mit} to solve the corresponding signal model in \eqref{eq:linear_regression} and to recover the values corresponding to correlations $\rho_{yx_i},~i\in\{1,2,3\}$. 
Hence, we can completely characterize the entries in the matrix $\mathbf{A_B}$ (up to sign), and the variances regarding additive Gaussian noise variables $Z_1$, $Z_2$, and $Z_3$.
However, one can only partially infer the sign, by observing the sign values corresponding to each $\rho_{x_ix_j},~i,j\in\{1,2,3\}$. 
In particular, such an approach leads us into two equivalent cases with sign inputs $b^{(1)}$ or $-b^{(1)}$, with the latter obtained by flipping the signs of all correlations $\rho_{yx_i},~i\in\{1,2,3\}$.
From \cite{correlation}, we know for the channel shown in Figure \ref{fig:broadcast}, we should have $\rho_{x_1x_2}\rho_{x_1x_3}\rho_{x_2x_3}>0$. Hence, there are totally two cases for $\rho_{x_ix_j},~i\neq j,~i,j\in\{1,2,3\}$ based on such constraint; either all of them are positive, or two of them are negative and the third one is positive.
As a result, one can classify the sign singularity for the broadcast channel shown in Figure \ref{fig:broadcast} into four groups, each consisting of two instances corresponding to $b^{(1)}$ or $-b^{(1)}$. 
For example, suppose we are given enough samples to infer the latent structure shown in Figure \ref{fig:broadcast}, in which all the pairwise correlations $\rho_{x_1x_2}$, $\rho_{x_1x_3}$, and $\rho_{x_2x_3}$ are derived as positive values. However, we cannot further decide on whether all pairwise correlations $\rho_{yx_i},~i\in\{1,2,3\}$ are positive, or all of them are negative, i.e., the ambiguity to choose between $b^{(1)}$ or $-b^{(1)}$.
Figure \ref{fig:sign} shows each group consisting of two Gaussian trees, in which the inferred correlation signs for $\rho_{yx_i},~i\in\{1,2,3\}$ are based on the signs of $\rho_{x_ix_j},~i,j\in\{1,2,3\}$.

Thus one may see that given only the observable vector $\mathbf{X}$, there is not enough information to distinguish $\mathbf{b}$ from $\mathbf{-b}$, hence the correlations sign information is partially lost. 
In Theorem \ref{thm:B}, whose proof can be found in Appendix \ref{app:B}, we characterize the size and dependency relations of sign vectors for any general minimal latent Gaussian tree.

\begin{thm} \label{thm:B}
{\it
$(1)$ The correlation values $\rho_{yx_i}$ in regard to the outputs $X_i$ that are connected to a single input, say $Y$, share an equivalent sign class, i.e., they either all belong to $B=b$ or $B=-b$.

$(2)$ Given the cardinality of input vector $\mathbf{Y}=\{Y_1,Y_2,...,Y_k\}$ is $k$, then there are totally $2^k$ minimal Gaussian trees with isomorphic structures, but with different correlation signs that induce the same joint density of the outputs, i.e., equal $p_{\mathbf{X}}(\mathbf{x})$.
}
\end{thm}

For example, in a Gaussian tree shown in Figure \ref{fig:broadcast}, there is only one hidden node $Y^{(1)}$, and we already know by previous discussions that there are two latent Gaussian trees with different sign values for $B^{(1)}$, which induce the same output joint density $p_{\mathbf{X}}(\mathbf{x})$.
In more general cases the problem of assigning correlation sign variables is more subtle, where we clarify the approach using two examples, next.

\begin{figure}[h!]
    \begin{subfigure}[h]{\columnwidth}
    \centering
        \includegraphics[width=0.35\columnwidth]{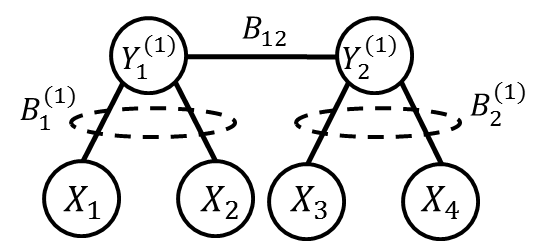}
        \caption{}
        \label{fig:B1}
    \end{subfigure}
    
    ~ 
    \begin{subfigure}[h]{\columnwidth}
    \centering
        \includegraphics[width=0.6\columnwidth]{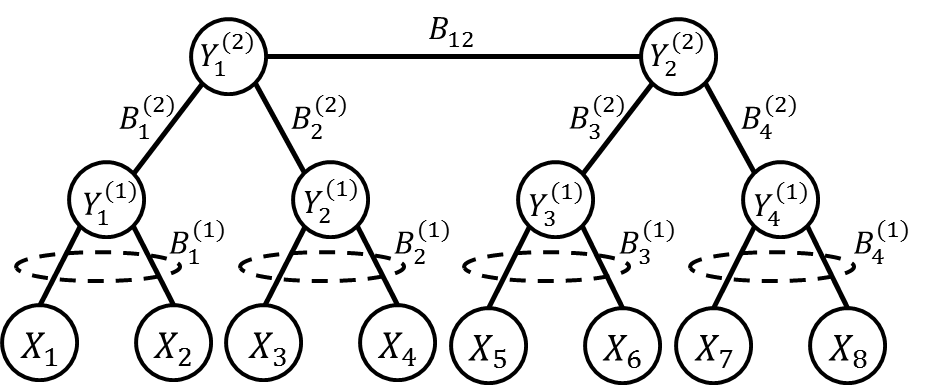}
        \caption{}
        \label{fig:B2}
    \end{subfigure}
    \caption{Two possible cases to demonstrate the dependency relations of sign variables: (a) with two hidden inputs, and (b) with $4$ hidden inputs at two layers}\label{fig:B}
\end{figure}
In a Gaussian tree shown in Figure \ref{fig:B1} there are two hidden nodes $Y_1$ and $Y_2$. By Theorem \ref{thm:B}, we know that there are $4$ Gaussian trees with sign ambiguity. 
Also, from the first part in Theorem \ref{thm:B} we may introduce $B^{(1)}_1$ to capture the correlation signs $\rho_{x_1y_1}$ and $\rho_{x_2y_1}$, and $B^{(1)}_2$ for the correlation signs $\rho_{x_3y_2}$ and $\rho_{x_4y_2}$.
We introduce $B_{12}$ as the sign of $\rho_{y_1y_2}$.
Note that the link between the variables $Y_1$ and $Y_2$ are in both groups with common correlation sign, so we anticipate that $B_{12}$ should be dependent on both $B^{(1)}_1$ and $B^{(1)}_2$.
Since we need to maintain the correlation signs regarding $\rho_{x_ix_j},~i\in\{1,2\},~j\in\{3,4\}$, hence the product $B^{(1)}_1B^{(1)}_2B_{12}$ should maintain its sign. 
Thus, we have $B_{12}=B^{(1)}_1B^{(1)}_2$, so $B_{12}$ is completely determined given $B^{(1)}_1$ and $B^{(1)}_2$. 
Next, consider the Gaussian tree shown in Figure \ref{fig:B2}, in which there are six hidden inputs. 
Similar to the previous case, we can show that $B^{(1)}_1B^{(1)}_2=B^{(2)}_1B^{(2)}_2$, $B^{(1)}_3B^{(1)}_4=B^{(2)}_3B^{(2)}_4$, and $B_{12}=B^{(1)}_1B^{(2)}_1B^{(2)}_4B^{(1)}_4$. Since there are nine sign values and three equality constraints, we have six free binary variables to represent all equivalent Gaussian trees under the constraint of sign ambiguity issues.

\section{Main Results} \label{sec:main}
\subsection{Maximum achievable rate region to generate the output $\mathbf{X}$}
In \cite{wyner}, a \textit{common information} of variables in $\mathbf{X}$ is defined to be the minimum rate among all possible sources, through which we can generate the outputs $\mathbf{X}$ with joint density $\hat{p}_{\mathbf{X}}(\mathbf{x})$ that is asymptotically close (measured by KL-distance) to the true joint density $p_{\mathbf{X}}(\mathbf{x})$.
Let us define $\mathbf{\tilde{Y}}=\{\mathbf{Y},\mathbf{B}\}$, then the formalized problem has the following form:
\begin{align} \label{eq:common_info}
C(\mathbf{X})&=\inf_{p_\mathbf{\tilde{Y}}(\mathbf{\tilde{y}})} I(\mathbf{X};\mathbf{\tilde{Y}}),~s.t.,\notag\\
&p_{\mathbf{X},\mathbf{\tilde{Y}}}(\mathbf{x},\mathbf{\tilde{y}})~induces~a~minimal~Gaussian~tree\notag\\
&X_i\perp X_j|\mathbf{\tilde{Y}}\notag\\
&\Sigma_{\tilde{y}\in\mathbf{\tilde{Y}}} p(\mathbf{x},\mathbf{\tilde{y}})=p_{\mathbf{X}}(\mathbf{x})
\end{align}
Note that all of the mutual information values should be evaluated under a given Gaussian tree $G_T(V,E,W)$. 
However, for simplicity we drop this notation in their expressions.
In this setting, by Theorem \ref{thm:fixed}, whose proof can be found in Appendix \ref{app:fixed}, we show that  regardless of the underlying Gaussian tree structure, there is no room to minimize $I(\mathbf{X};\tilde{\mathbf{Y}})$.

\begin{thm} \label{thm:fixed}
{\it
Given $p_{\mathbf{X}}(x)\sim~N(0,\Sigma_\mathbf{x})$ and the settings in \eqref{eq:common_info}, the mutual information $I(\mathbf{X};\mathbf{\tilde{Y}})$ is only a function of $\Sigma_\mathbf{x}$ and if the observable nodes are only leaf nodes, the mutual information is given by,
\begin{align}
I(\mathbf{X};\mathbf{\tilde{Y}})=\dfrac{1}{2}\log\dfrac{|\Sigma_\mathbf{x}|}{\prod_{i=1}^n (1-\dfrac{\rho_{x_ix_{j_i}}\rho_{x_ix_{k_i}}}{\rho_{x_{j_i}x_{k_i}}})}
\end{align}
\noindent where for each $X_i$, we choose two other nodes $X_{j_i}$, $X_{k_i}$, where all three of them are connected to each other through $Y_{X_i}$ (i.e., one of their common ancestors), which is one of the hidden variables adjacent to $X_i$.
}
\end{thm}
\begin{rmk}
Intuitively, given $\Sigma_x$ and any three outputs that have a common latent variable as their input, the correlation values between each output and the input is fixed, since varying one correlation results in varying the other correlations in the same direction, hence making the pairwise correlation between the other outputs change, which is impossible. 
Also, as we may observe from Theorem \ref{thm:fixed}, given $X_i$ we may end up with several options for $X_{j_i}$ and $X_{k_i}$.
However, it can be shown that in a subspace of correlations corresponding to latent Gaussian trees \cite{correlation}, all those distinct options result in a same value for the term $\rho_{x_ix_{j_i}}\rho_{x_ix_{k_i}}/\rho_{x_{j_i}x_{k_i}}$.
\end{rmk}
\begin{rmk}
By \eqref{eq:linear_regression}, one may see that changing the sign vector $\mathbf{B}$ does not influence the output vector $\mathbf{X}$, hence we can show that $I(\mathbf{X};\mathbf{B})=0$.
For example, consider the star model in Figure \ref{fig:broadcast}. 
Since the input elements $Y_i\in\mathbf{Y}$ have zero means, so the conditional density becomes as $p_{\mathbf{X}|\mathbf{B}}(\mathbf{x}|\mathbf{b})\sim N(\mathbf{0},\Sigma_{\mathbf{x}|\mathbf{b}})$.
By varying $B_i\in\mathbf{B}$ we are just flipping the conditional density around the origin, which does not change the corresponding conditional entropy $h(\mathbf{X}|\mathbf{B})$, hence making $\mathbf{X}$ and $\mathbf{B}$ independent.
From the equality $I(\mathbf{X};\mathbf{\tilde{Y}})=I(\mathbf{X};\mathbf{B})+I(\mathbf{X};\mathbf{Y}|\mathbf{B})$ and above arguments, we know $I(\mathbf{X};\mathbf{\tilde{Y}})=I(\mathbf{X};\mathbf{Y}|\mathbf{B})$, which explains why previous learning algorithms have neglected the sign ambiguity carried by the sign vector $\mathbf{B}$ by assuming that the sign information is given and aiming to infer the model.
\end{rmk}
\begin{rmk}
One may easily deduce the following,
\begin{align} \label{eq:equality}
I(\mathbf{X};\mathbf{\tilde{Y}})=I(\mathbf{X};\mathbf{Y},\mathbf{B})=I(\mathbf{X};\mathbf{Y}) + I(\mathbf{X};\mathbf{B}|\mathbf{Y})
\end{align}
The result in Theorem \ref{thm:fixed} combined with \eqref{eq:equality}, suggests that by minimizing $I(\mathbf{X};\mathbf{Y})$, one may eventually maximize $I(\mathbf{X};\mathbf{B}|\mathbf{Y})$ 
, i.e., quantifying the maximum amount of information loss on the sign input $\mathbf{B}$.
\end{rmk}

\subsection{Synthesis of the Gaussian output vector $\mathbf{X}$ given $p_{\mathbf{X}}(\mathbf{x})$}
In this section we provide mathematical formulations to address the following fundamental problem: using channel inputs $\mathbf{Y}$ and $\mathbf{B}$, what are the rate conditions under which we can synthesize the Gaussian channel output $\mathbf{X}$, with a given $p_{\mathbf{X}}(\mathbf{x})$.
We propose an encoding scheme, as well as the corresponding bounds on achievable rate tuples.

Suppose we transmit input messages through $N$ channel uses, in which $t\in\{1,2,...,N\}$ denotes the time index.
We define $\vec{Y}^{(l)}_{t}[i]$ to be the $t$-th symbol of the $i$-th codeword, with $i\in\{1,2,...,M_{Y^{(l)}}\}$ where $M_{Y^{(l)}}=2^{NR_{Y^{(l)}}}$ is the codebook cardinality, transmitted from the existing $k_l$ sources at layer $l$.
Here, we define the source to be at layer $l$, if the shortest path from source to output passes through $l$ links.
Also, we assume there are $k_l$ sources $Y^{(l)}_j$ present at the $l$-th layer, and the channel has $L$ layers.
We can similarly define $\vec{B}^{(l)}_{t}[k]$ to be the $t$-th symbol of the $k$-th codeword, with $k\in\{1,2,...,M_{B^{(l)}}\}$ where $M_{B^{(l)}}=2^{NR_{B^{(l)}}}$ is the codebook cardinality, transmitted from the existing $k_l$ sources at layer $l$.
For \textit{sufficiently} large rates $R_{\mathbf{Y}}=[R_{Y^{(1)}},R_{Y^{(2)}},...,R_{Y^{(L)}}]$ and $R_{\mathbf{B}}=[R_{B^{(1)}},R_{B^{(2)}},...,R_{B^{(L)}}]$ and as $N$ grows the output density of synthesized channel converges to $p_{\mathbf{X}^N(\mathbf{x}^N)}$, i.e., $N$ i.i.d realization of the given output density $p_{\mathbf{X}}(\mathbf{x})$.
In other words, the average total variation between the two joint densities vanishes as $N$ grows \cite{cuff},
\begin{align} \label{eq:TV}
\lim_{N\rightarrow\infty} E||q(\mathbf{x}_1,...,\mathbf{x}_N)-\prod_{t=1}^N p_{\mathbf{X}_t}(\mathbf{x}_t)||_{TV}\rightarrow 0
\end{align}
\noindent where $q(\mathbf{x}_1,...,\mathbf{x}_N)$ is the synthesized channel output, and $E||.||_{TV}$, represents the average total variation.
In this situation, we say that the rates $(R_{\mathbf{Y}},R_{\mathbf{B}})$ are \textit{achievable} \cite{cuff}.
 For example, for the channel shown in Figure \ref{fig:broadcast} we may compute the synthesized output as,
\begin{align}
&q(\mathbf{x}_1,...,\mathbf{x}_N)=\notag\\
&\dfrac{1}{M_{\mathbf{B}}}\dfrac{1}{M_\mathbf{Y}}\sum_{i=1}^{M_\mathbf{Y}}\sum_{k=1}^{M_\mathbf{B}}\prod_{t=1}^N p_{\mathbf{X}|\mathbf{Y},\mathbf{B}}(\mathbf{x}_t|\mathbf{y}_t[i]\mathbf{b}_t[k])
\end{align}
\noindent where $M_{\mathbf{B}}=2^{NR_{\mathbf{B}}}$ and $M_{\mathbf{Y}}=2^{NR_{\mathbf{Y}}}$ are the total number of input messages for sources $\mathbf{B}$ and $\mathbf{Y}$, respectively. Also, the distribution $p_{\mathbf{X}|\mathbf{Y},\mathbf{B}}(\mathbf{x}_t|\mathbf{y}_t[i]\mathbf{b}_t[k])$ represents each channel use $t$ for corresponding input messages, and can be computed via signal model in \eqref{eq:linear_regression}. 

In the following sections, we provide three case studies through which we obtain achievable rate regions to synthesize the output statistics $p_{\mathbf{X}}(\mathbf{x})$. 
Here, for simplicity of notation, we drop the symbol index and use $Y^{(l)}_{t}$ and $B^{(l)}_{t}$ instead of $\vec{Y}^{(l)}_{t}[i]$ and $\vec{B}^{(l)}_{t}[k]$, respectively, since they can be understood from the context.
\subsubsection{Channel Synthesis for the Star Model}

Consider a broadcast channel with Gaussian source $Y^{(1)}$ and sign input $B^{(1)}$, with corresponding output vector $\mathbf{X}=[X_1,X_2,...,X_n]$. This can be modeled as
\begin{align} \label{eq:regression_star}
\begin{bmatrix}
X_{1,t}\\
X_{2,t}\\
\vdots\\
X_{n,t}
\end{bmatrix}
=
\begin{bmatrix}
\alpha_1\\
\alpha_2\\
\vdots\\
\alpha_n
\end{bmatrix}
B^{(1)}_t
Y^{(1)}_t
+
\begin{bmatrix}
Z_{1,t}\\
Z_{2,t}\\
\vdots\\
Z_{n,t}
\end{bmatrix}
,~t\in\{1,2,...,N\}
\end{align}

A special case for such broadcast channel is shown in Figure \ref{fig:broadcast}, where the channel has only three outputs $X_1$, $X_2$, and $X_3$.
Due to soft covering lemma and the results in \cite{cuff} we have the following Theorem, whose proof can be found in Appendix \ref{app:achievability1}.
\begin{thm} \label{thm:achievability1}
{\it
For the broadcast channel characterized by \eqref{eq:regression_star}, the following rates are achievable,
\begin{align} \label{eq:conditions}
R_{Y^{(1)}}+R_{B^{(1)}}\geq I(\mathbf{X};Y^{(1)},B^{(1)})\notag\\
R_{Y^{(1)}}\geq I(\mathbf{X};Y^{(1)})
\end{align}
}
\end{thm}

Note that the sum of the rates $R_{Y^{(1)}}+R_{B^{(1)}}$ is lower bounded by $I(\mathbf{X};Y^{(1)},B^{(1)})$, which by Theorem \ref{thm:fixed} is fixed.
However, the minimum rate for $R_{Y^{(1)}}$ is achieved by $min_{Y^{(1)}} I(\mathbf{X};Y^{(1)})$.
In the following Theorem we prove that the optimal solution occurs when $B^{(1)}$ is uniformly distributed.
\begin{thm} \label{thm:broadcast}
{\it
The optimal solution to the optimization problem $\pi^*=arg\max_{\pi_1\in [0,1]} I(\mathbf{X};B^{(1)}|Y^{(1)})$ is $\pi^*=1/2$.
}
\end{thm}

\begin{proof}
The proof relies on the results shown in \cite{capacity}. One can show that given $Y^{(1)}=y$, the broadcast model in \eqref{eq:regression_star} becomes a bipolar signaling scheme with $\mathbf{S}_1=[\alpha_1y,\alpha_2y,...,\alpha_ny]'$ and $\mathbf{S_2}=-\mathbf{S_1}$.
Now, simply by putting $\mathbf{T}=-\mathbf{I}_n$, where $\mathbf{T}$ is an orthonormal matrix and using $\mathbf{S_2}=\mathbf{TS_1}$, one can map the signals to each other.
Also, we may normalize the noise variances to satisfy all the constraints shown in \cite{capacity}. This can be simply done by introducing $n\times n$ diagonal matrix $\mathbf{M}$ with $m_{ii}=1/\sigma_{z_i}$, through which the new signal model becomes $\mathbf{X}'=\mathbf{MX}=\mathbf{MS_i}+\mathbf{MZ},~i\in\{1,2\}$, where $Z'_i\in\mathbf{Z}'=\mathbf{MZ}$ are noises with unit variance.
Since such model is \textit{circular symmetric} \cite{capacity}, hence, $\pi^*=1/2$.
\end{proof}

\subsubsection{Channel Synthesis with $4$ outputs and $2$ inputs}

Consider the channel shown in Figure \ref{fig:B1}. In this case, we are given two hidden inputs $Y_1^{(1)}$ and $Y_2^{(1)}$, and by previous arguments we know $\mathbf{B}^{(1)}=\{B_1^{(1)},B_2^{(1)},B_{12}\}$ with $B_{12}=B_1^{(1)}B_2^{(1)}$, completely determined by $B_1^{(1)}$ and $B_2^{(1)}$, which may act independently. 
We may write,
\begin{align}
\begin{bmatrix}
X_{k,t}\\
X_{l,t}
\end{bmatrix}
=
\begin{bmatrix}
\alpha_{j,k}\\
\alpha_{j,l}
\end{bmatrix}
B^{(1)}_{j,t}
Y^{(1)}_{j,t}
+
\begin{bmatrix}
Z_{k,t}\\
Z_{l,t}\\
\end{bmatrix}
,~t\in\{1,2,...,N\}
\end{align}
\noindent where for $j\in\{1,2\}$, $(k,l)\in\{(1,2),(3,4)\}$, which shows each group of outputs corresponding to each of the inputs.
Here, two inputs $Y_1^{(1)}$ and $Y_2^{(1)}$ are dependent and their pairwise correlation can be computed via $E[Y_1^{(1)}Y_2^{(1)}|\mathbf{B}^{(1)}]=\gamma_{12}B_{12}=\gamma_{12}B_1^{(1)}B_2^{(1)}$, in which $\gamma_{12}$ determines the degree of correlation and is learned by certain inference algorithms, e.g., RG or CLRG \cite{mit}.
Note that the dependency relation of symbols $Y_{1,t}^{(1)}$ and $Y_{2,t}^{(1)}$ follows a Gaussian mixture model, since their covariance is a function of binary inputs $B_{1,t}^{(1)}$ and $B_{2,t}^{(1)}$.
But, note that in a given codebook consisting of $M_{\mathbf{Y}^{(1)}}\times M_{\mathbf{B}^{(1)}}$ codewords, for each realization of $\mathbf{b}_{3,t}^{(1)}=\mathbf{b}_{1,t}^{(1)}\mathbf{b}_{2,t}^{(1)}$ the joint density of $\mathbf{Y}_{t}^{(1)}$ is Gaussian. 
Hence, one may divide the codebook $\mathbb{C}$ into two parts $\mathbb{S}_i,~i\in\{1,2\}$, in which each part follows a specific Gaussian density with covariance values $E[Y_{k,t}^{(1)}Y_{l,t}^{(1)}|\mathbf{b}^{(1)}]=\gamma_{12}b_{1,t}^{(1)}b_{2,t}^{(1)}$.
Note that such sub-block encoding guarantees the independence between the synthesized output vector and the sign input vector $\mathbf{B}$.
The achievable region can be obtained from \eqref{eq:conditions}, and by replacing $Y^{(1)}$ with $\{Y^{(1)}_1,Y^{(1)}_2\}$ and $B^{(1)}$ with $\{B^{(1)}_1,B^{(1)}_2\}$. The achievability of the rates can be shown using similar steps as taken in the proof of Theorem \ref{thm:achievability1}.
 
In the following Lemma, whose proof can be found in Appendix \ref{app:optimal_dumbbell}, we showed that the optimal solution $(\pi^*_1,\pi^*_2)$ to $arg\max_{\pi_1,\pi_2} I(\mathbf{X};\mathbf{B}|\mathbf{Y})$ is at $(1/2,1/2)$.
\begin{lem} \label{lem:optimal_dumbbell}
{\it
For the channel shown in Figure \ref{fig:B1}, we have,

$(1)$ $I(\mathbf{X};\mathbf{B}|\mathbf{Y})=I(X_1,X_2;B_1^{(1)}|Y_1^{(1)}Y_2^{(1)})+I(X_3,X_4;B_2^{(1)}|Y_1^{(1)}Y_2^{(1)})$.

$(2)$ The optimal solution for the maximization problem $arg\max_{\pi_1,\pi_2} I(\mathbf{X};\mathbf{B}|\mathbf{Y})$ happens at $\pi^*_1=\pi^*_2=1/2$.
}
\end{lem}

Intuitively, for the first part in Lemma \ref{lem:optimal_dumbbell}, we may divide the structure shown in Figure \ref{fig:B1} into two substructures each similar to the star topology shown in Figure \ref{fig:broadcast} (but with only two outputs). Hence, we may use the results in Theorem \ref{thm:broadcast} to prove the second part of the lemma.

\subsubsection{Multi-Layered Channel Synthesis}
\begin{figure} [h!]
\centering 
\includegraphics[scale=0.6]{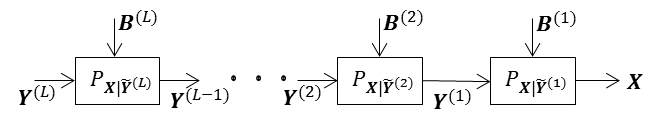}
\caption{Multi-layered output synthesis\label{fig:layered}} 
\end{figure}

Here, we address those channels with multi-layer inputs. 
Figure \ref{fig:layered} shows the general encoding scheme to be used to synthesize the output vector.
At each layer $i$, we define $\tilde{\mathbf{Y}}^{(i)}=\{\mathbf{Y}^{(i)},\mathbf{B}^{(i)}\}$ to be the combination of input vectors.
This situation is a little more subtle than the previous single-layered cases, since we need to be cautious on defining the rate regions.

To clarify, consider the case shown in Figure \ref{fig:B2}, in which the Gaussian tree has two layers of inputs. Similar as previous cases we may write the encoding scheme, through which we can write the pairwise covariance between inputs at the first layer as $E[Y_{k,t}^{(1)}Y_{l,t}^{(1)}|\mathbf{B}^{(1)}]=\gamma_{kl}B_{k,t}^{(1)}B_{l,t}^{(1)}$, in which $k\neq l\in\{1,2,3,4\}$. 
By the previous example, we know that the input vector $\mathbf{Y}_{t}^{(1)}$ becomes Gaussian for each realization of $\mathbf{B}_{t}^{(1)}=\{\mathbf{b}_{1,t}^{(1)},\mathbf{b}_{2,t}^{(1)},\mathbf{b}_{3,t}^{(1)},\mathbf{b}_{4,t}^{(1)}\}$. 
Hence, one may divide the codebook $\mathbb{C}$ into $2^4=16$ parts $\mathbb{S}_i,~i\in\{1,2,...,16\}$, in which each part follows a specific Gaussian density with covariance values $E[Y_{k,t}^{(1)}Y_{l,t}^{(1)}|\mathbf{b}^{(1)}]=\gamma_{kl}b_{k,t}^{(1)}b_{l,t}^{(1)},~k\neq l\in\{1,2,3,4\}$.
Now, for each subset, at the second layer we are dealing with the case shown in Figure \ref{fig:B1}, which has been resolved.
Thus, the lower bound on the possible rates in the second layer are, 
\begin{align}
&R_{\mathbf{Y}^{(2)}|\mathbf{B}^{(1)}}\geq I(\mathbf{Y}^{(1)};\mathbf{Y}^{(2)}|\mathbf{B}^{(1)})\notag\\
&R_{\mathbf{Y}^{(2)}|\mathbf{B}^{(1)}}+R_{\mathbf{B}^{(2)}|\mathbf{B}^{(1)}}\geq I(\mathbf{Y}^{(1)};\mathbf{Y}^{(2)},\mathbf{B}^{(2)}|\mathbf{B}^{(1)})
\end{align}
This is due to the fact that we compute subsets of codebook for each realization of $\mathbf{B}^{(1)}$.
Hence, in general the output at the $l$-th layer $\mathbf{Y}^{(l)}$ is synthesized by $\mathbf{Y}^{(l+1)}$ and $\mathbf{B}^{(l+1)}$, which are at layer $l+1$.
Therefore, we only need Gaussian sources at the top layer $L$ and Bernoulli sources $B^{(l)}$ for each layer $l$ to gradually synthesize the output that is close enough to the true observable output, measured by total variation.

\section{Conclusion} \label{sec:conclusion}
We studied the sign singularity of current latent Gaussian trees, which results in partially losing the correlation signs information in the recovered model.
We then formulated a Gaussian synthesis problem through layered forwarding channels to synthesize the observed data.
Then we deduced an interesting conclusion under which maximizing the achievable rate region also resulted in quantifying the maximum amount of lost information on pairwise correlation signs.
Through three different case studies we found the achievable rate regions to correctly synthesize the Gaussian output, satisfying specific set of constraints.
In the future, we aim to investigate those channels with arbitrary number of layers and having one or several outputs acting as internal variables for the underlying latent Gaussian tree structure.

\bibliography{reference}
\bibliographystyle{IEEEtran}

\appendices

\section{Proof of Theorem \ref{thm:B}} \label{app:B}

\begin{figure} [h!]
\centering
\includegraphics[scale=0.5]{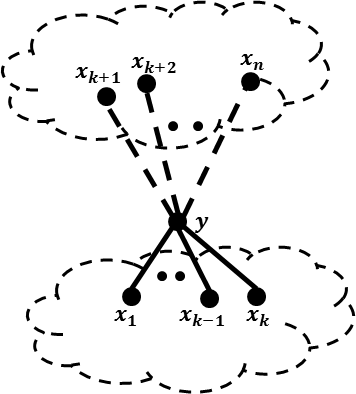} 
\caption{Neighborhood of hidden variable $y$}\label{fig:neighbor} 
\end{figure}

First, let's prove the first part.
Consider the case in Figure \ref{fig:neighbor}. The hidden node $y$, has $k$ observable neighbors $\{x_1,...,x_k\}$, while it is connected through two or more edges to other observable nodes $\{x_{k+1},...,x_n\}$. 
Given only observable covariance matrix $\Sigma_x$, we can compute the empirical pairwise covariance values, hence all $\rho_{x_ixj}$ are fixed.

Without loss of generality, suppose we flip the sign of $\rho_{x_1y}$. 
In order to maintain the same covariance matrix $\Sigma_x$, the sign of all $\rho_{x_jy}$, $j\in\{2,...,k\}$ should be flipped. This is easy to see because for all $j\in\{2,...,k\}$, we know $\rho_{x_1x_j}=\rho_{x_1y}\rho_{x_jy}$, is fixed.
Also, the sign of all pairwise covariance values between $y$ and $x_i$, for all $i\in\{k+1,...,n\}$ should be flipped. The same argument as the previous case can be used. However, in this case, all we know is that odd number of sign-flips for the edge-weights between each $y$ and $x_i$ should happen.
Using the above arguments, we can see that all $\rho_{x_jy}$ for $j\in\{1,...,k\}$ maintain their signs, or otherwise all of their signs should be flipped.

\begin{figure} [h!]
\centering
\includegraphics[scale=0.5]{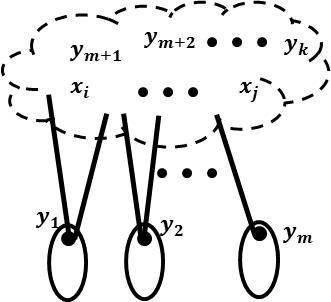} 
\caption{Figure illustrating the inductive proof}\label{fig:induction} 
\end{figure}

For the second part, 
We inductively show that given a minimal latent tree, with $n$ observable $x_1,...,x_n$ and with $k$ hidden nodes $y_1,...,y_k$, we can find $2^k$ latent trees with different edge-signs that induce the same $\Sigma_x$.
This is already shown for the star tree shown in Figure \ref{fig:broadcast}.
Suppose such claim holds for all Gaussian trees with $k'<k$ latent nodes. 
Consider an arbitrary latent tree with $k$ hidden nodes and $n$ observable.
Some of these hidden nodes certainly have leaf observable neighbors, which we group them together.
Now, note that the problem of finding equivalent sign permutations in this tree can be translated into a problem with smaller tree: Delete all of those leaf observable groups, and treat their hidden parent $y_i$ as their representative. Suppose there are $m$ hidden nodes $\{y_1,...,y_m\}$, which can represent each of these groups. This case is illustrated in Figure \ref{fig:induction}. 
Note, as depicted by this Figure, the internal observables as well as those leaf observables directly connected to them remain intact.
By replacing all of these groups with a single node $y_i$, for all $i\in\{1,2,...,m\}$, we obtain a smaller tree. Now, we can simply assume that all $y_1,...,y_m$ are observable and their pairwise covariance values are determined. Hence, this tree only has $k-m$ remaining hidden nodes, so due to inductive step it has $2^{k-m}$ possible equivalent trees with different edge-signs.

\begin{figure} [h!]
\centering
\includegraphics[scale=0.5]{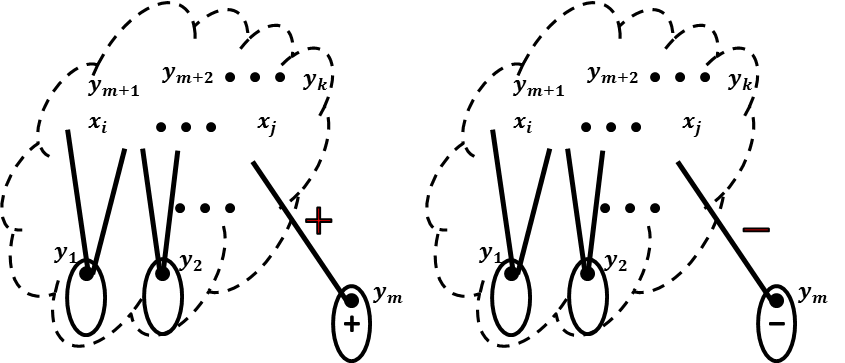} 
\caption{Obtaining $m$-th step from $m-1$-th step}\label{fig:mth} 
\end{figure}

All its remaining is to show that by adding back those $m$ groups of observable, we obtain the claimed result. Add back two groups corresponding to $y_1$ and $y_2$. Now, $y_1$ and $y_2$ can be regarded as hidden nodes, so now there are $k-m+2$ hidden nodes, which due to inductive step has $2^{k-m+2}$ equivalent representations of edge-weights. This can be shown up to $m-1$-th step by adding back the groups for $y_1,...,y_{m-1}$ nodes, and having a size of $k-1$ nodes, and again due to induction having $2^{k-1}$ equivalent sign combinations.
By adding back the $m$-th group, we can obtain two equivalent classes: $b^{(m)}$ or $-b^{(m)}$, where $b^{(m)}$ shows the sign value of the $m$-th group.  
This is shown in Figure \ref{fig:mth}
Hence, we obtain $2\times 2^{k-1}=2^k$ edge-signs.

This completes the proof.

\section{Proof of Theorem \ref{thm:fixed}} \label{app:fixed}

Let's first show that the mutual information $I(\mathbf{X},\tilde{\mathbf{Y}})$ given $\Sigma_{\mathbf{x}}$ is only a function of pairwise correlations $\rho_{x_ix_j}$, for all $x_i,x_j\in\mathbf{X}$.
In a latent Gaussian tree, three cases may happen: The edges can be between two observable, an observable and a latent node, or between two latent nodes.

$(1)$ $x_i$ and $x_j$ are either adjacent or they are connected only through several observables.
In this case, since all the pairwise correlations along the path are determined given $\Sigma_{\mathbf{x}}$, so the correlation values are fixed.

$(2)$ $x_i$ and $x_j$ are not adjacent and there is at least one hidden node, e.g., $y_1$ connecting them. First, suppose $y_1$ and $x_i$ are adjacent. Since, we assume the tree is minimal, so there should be at least another observable $x_k$ that is connected (but not necessarily adjacent) to $y_1$. Hence, $y_1$ acts as a common ancestor to $x_i$, $x_j$, and $x_k$. By changing $\rho_{x_iy_1}$ to another value $\rho'_{x_iy_1}$, by equation $\rho_{x_ix_j}=\rho_{x_iy_1}\rho_{x_jy_1}$ we have to change $\rho_{x_jy_1}$ to $\rho'_{x_jy_1}=\dfrac{\rho_{x_iy_1}}{\rho'_{x_iy_1}}\rho_{x_jy_1}$. Similarly, by equality $\rho_{x_ix_k}=\rho_{x_iy_1}\rho_{x_ky_1}$, we know $\rho'_{x_ky_1}=\dfrac{\rho_{x_iy_1}}{\rho'_{x_iy_1}}\rho_{x_ky_1}$. However, by another equality $\rho_{x_jx_k}=\rho_{x_jy_1}\rho_{x_ky_1}$, we deduce $\rho'_{x_ky_1}=\dfrac{\rho_{x_jy_1}}{\rho'_{x_jy_1}}\rho_{x_ky_1}$. The obtained correlation $\rho'_{x_ky_1}$ should have the same value in both equations, hence, we deduce the equality $\dfrac{\rho_{x_iy_1}}{\rho'_{x_iy_1}}=\dfrac{\rho_{x_jy_1}}{\rho'_{x_jy_1}}$.
On the other hand, from $\rho_{x_ix_j}=\rho_{x_iy_1}\rho_{x_jy_1}$, we have $\dfrac{\rho_{x_iy_1}}{\rho'_{x_iy_1}}=\dfrac{\rho'_{x_jy_1}}{\rho_{x_jy_1}}$.
By these two equations we may conclude $\rho_{x_iy_1}=\rho'_{x_iy_1}$, a contradiction.
Hence, in this case, given $\Sigma_\mathbf{x}$ we cannot further vary the edge-weights.
Second, consider the case, where $x_i$ is connected to $y_1$ through several observables.
Then, instead of $x_i$, we can simply consider the observable that is adjacent to $y_1$, say, $x'_i$ and follows the previous steps to obtain the result.
Hence, in general if three nodes are connected to each other through separate paths and have a common ancestor $y_1$, then the pairwise correlations between the hidden nodes and each of the observables remain fixed.

$(3)$  Consider two adjacent latent nodes $y_1$ and $y_2$. By minimality assumption and having a tree structure, we can argue that there are at least two observable for each of the latent nodes that share a common latent parent. Let's assign $x_i$ and $x_j$ to a common ancestor $y_1$ while $x_k$ and $x_k$ are descendant to $y_2$. Considering $x_i$, $x_j$, and $x_k$, who share a common parent $y_1$ ($x_k$ is connected to $y_1$ through $y_2$), using arguments on case $(2)$, we conclude that $\rho_{x_iy_1}$ and $\rho_{x_jy_1}$ should be fixed.
Similarly, we can consider $x_i$, $x_k$, and $x_l$ to show that $\rho_{x_ky_1}$ and $\rho_{x_ly_1}$ are fixed.
Now, by considering any observable pair that go through both $y_1$ and $y_2$ the result follows. For example, considering $\rho_{x_ix_k}=\rho_{x_iy_1}\rho_{y_1y_2}\rho_{x_ky_1}$, we can see that since given $\rho_{x_ix_k}$, both $\rho_{x_iy_1}$ and $\rho_{x_ky_1}$ are determined, so $\rho_{y_1y_2}$ should be determined as well.
This completes the first part of the proof.

Second, note that one may easily show that $I(\mathbf{X},\tilde{\mathbf{Y}})=1/2\log\dfrac{|\Sigma_\mathbf{x}||\Sigma_{\tilde{\mathbf{y}}}|}{|\Sigma_{\mathbf{x}\tilde{\mathbf{y}}}|}$. Now, since $p_{\mathbf{X,\tilde{Y}}}$ induces a latent Gaussian tree and $p_{\tilde{\mathbf{Y}}}$ is its marginalized density after summing out the random vector $\mathbf{X}$. By \cite{arxiv}, we know that $|\Sigma_{\mathbf{X},\tilde{\mathbf{Y}}}|=\prod_{(i,j)\in E} (1-\rho^2_{i,j})$, where $\rho_{i,j}$ are the pairwise correlations, between two adjacent variables (hidden or observable) in a latent Gaussian tree.
Now, since the observables are only leaves, by summing them out we end with another Gaussian tree consisting of only latent variables.
Thus, again by \cite{arxiv} we know $|\Sigma_{\tilde{\mathbf{Y}}}|=\prod_{(i,j)\in E_y} (1-\rho^2_{i,j})$, where $E'$ is the set of edges in the new Gaussian tree.
Observe that all the common terms of the form $(1-\rho^2_{y_iy_j})$, for some $(y_i,y_j)\in E$ will be canceled out with the terms in $|\Sigma_{\tilde{\mathbf{Y}}}|$.
Hence, the mutual information has the following form $I(\mathbf{X},\tilde{\mathbf{Y}})=1/2\log\dfrac{|\Sigma|_\mathbf{X}}{\prod_{(x_i,y_j)\in E (1-\rho^2_{x_iy_j})}}$.
Now, to find each correlation value $\rho_{x_iy_j}$, for some $X_i$ and $Y_j$,
first consider the star model, with one hidden node, and three leaves, e.g., Figure \ref{fig:broadcast}. We can write: $\rho^2_{x_1y}=\dfrac{\rho_{x_1x_2}\rho_{x_1x_3}}{\rho_{x_2x_3}}$, $\rho^2_{x_2y}=\dfrac{\rho_{x_1x_2}\rho_{x_2x_3}}{\rho_{x_1x_3}}$, and $\rho^2_{x_3y}=\dfrac{\rho_{x_1x_3}\rho_{x_2x_3}}{\rho_{x_1x_2}}$. 
For a general structure, if we replace $1\leftarrow i$, $2\leftarrow j_i$, and $3\leftarrow k_i$, we conclude that $\rho^2_{x_iy_j}=\dfrac{\rho_{x_ix_{j_i}}\rho_{x_ix_{k_i}}}{\rho_{x_{j_i}x_{k_i}}}$, for any three distinct $i,~j_i$ and $k_i$. As it may seem, there are many equations for computing $\rho^2_{x_iy_j}$, which all of these expressions should be equal, i.e., the covariance matrix $\Sigma_x$ should be representable by a given latent tree model.

\section{Achievability Proof of Theorem \ref{thm:achievability1}} \label{app:achievability1}

The proof relies on the procedure taken in \cite{cuff}. Note that our encoding scheme should satisfy the following constraints,

\begin{tabular}{l l}
$1) X_i^N\perp X_j^N|\tilde{Y}~~ (i\neq j)$ & $4) |Y^{(1)}|= 2^{NR_{Y^{(1)}}}$\\
$2) \mathbf{X}^N\perp B^{(1)}$ & $5) |B^{(1)}|= 2^{NR_{B^{(1)}}}$\\
$3) \mathbf{X}^N~are~i.i.d\sim P_{\mathbf{X}}(\mathbf{x})$ & $6) ||q_{\mathbf{X}^N-\prod_{t=1}^N P_{\mathbf{X}(\mathbf{x}_t)}}||_{TV}<\epsilon$\\
\end{tabular}

\noindent where the first constraint is due to the underlying star model characterized in \eqref{eq:regression_star}. The second one is to capture the intrinsic ambiguity of the latent Gaussian tree (i.e., the star model) to capture the sign information. Condition $3)$ is due to the assumption of a given  Gaussian density $P_{\mathbf{X}}(\mathbf{x})\sim N(0,\Sigma_{\mathbf{X}})$ for the output vector. Conditions $4)$ and $5)$ are due to corresponding rates for each of the inputs $Y^{(1)}$ and $B^{(1)}$. And finally, condition $6)$ is the synthesis requirement to be satisfied.

\begin{figure} [h!]
\centering
\includegraphics[scale=0.6]{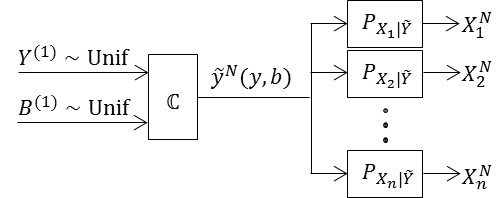} 
\caption{Construction of the joint density $\gamma_{\mathbf{X}^N,Y^{(1)},B^{(1)}}$}\label{fig:encoding} 
\end{figure}

First, we generate a codebook $\mathcal{C}$ of $\tilde{y}^N$ sequences, with indices $y\in C_Y=\{1,2,...,2^{NR_{Y^{(1)}}}\}$ and $b\in C_B=\{1,2,...,2^{NR_{B^{(1)}}}\}$ according to $\prod_{t=1}^N P_{\tilde{Y}}(\tilde{y}_t)$.
We construct the joint density $\gamma_{\mathbf{X}^N,Y^{(1)},B^{(1)}}$ as depicted by Figure \ref{fig:encoding},

The indices $y$ and $b$ are chosen independently and uniformly from the codebook $\mathcal{C}$.
As can be seen from Figure \ref{fig:encoding}, the channel $P_{\mathbf{X}|\tilde{Y}}$ is in fact consists of three independent channels $P_{X_i|\tilde{Y}},~i\in\{1,2,3\}$. The joint density has the following form,
\begin{align*}
\gamma_{\mathbf{X}^N,Y^{(1)},B^{(1)}}=\dfrac{1}{|C_Y||C_B|}[\prod_{t=1}^N P_{\mathbf{X}}(\mathbf{x}_t|\tilde{y}_t(y,b))]
\end{align*}

Note that $\gamma_{\mathbf{X}^N,Y^{(1)},B^{(1)}}$ already satisfies the constraints $1)$, $4)$, and $5)$ by construction.
Next, we need to show that it satisfies the constraint $6)$. The marginal density $\gamma_{\mathbf{x}^N}$ can be deduced by the following,
\begin{align*}
\gamma_{\mathbf{x}^N}=\dfrac{1}{|C_Y||C_B|}\sum_{y\in C_Y}\sum_{b\in C_B}[\prod_{t=1}^N P_{\mathbf{X}}(\mathbf{x}_t|\tilde{y}_t(y,b))]
\end{align*}

We know if $R_{B^{(1)}}+R_{Y^{(1)}} \geq I[\mathbf{X};\tilde{Y}]$, then by soft covering lemma \cite{cuff} we have,
\begin{align} \label{eq:6}
\lim_{n\rightarrow\infty} E||\gamma_{\mathbf{X}^n}-\prod P_{\mathbf{X}}||_{TV} = 0
\end{align}

\noindent which shows that $\gamma_{\mathbf{X}^N}$ satisfies constraint $6)$. For simplicity of notations we use $\prod P_{\mathbf{X}}$ instead of $\prod_{t=1}^N P_{\mathbf{X}}(\mathbf{x}_t)$, since it can be understood from the context.

Next, let's show that $\gamma_{\mathbf{X}^N}$, \textit{nearly} satisfies constraints $2)$ and satisfies $3)$. We need to show that as $N$ grows the synthesized density $\gamma_{\mathbf{X}^N,\mathbf{B}}$ approaches $\dfrac{1}{|C_B|}\prod P_{\mathbf{X}}$, in which the latter satisfies both $2)$ and $3)$. In particular, we need to show that the total variation 
$E||\gamma_{\mathbf{X}^N,B^{(1)}}-\dfrac{1}{|C_B|}\prod P_{\mathbf{X}}||$ vanishes as $N$ grows.
After taking several algebraic steps similar to the ones in \cite{cuff}, we should equivalently show that the following term vanishes, as $N\rightarrow\infty$,
\begin{align} \label{eq:nearly}
\dfrac{1}{|C_B|}\sum_{b\in C_B}E||\gamma_{\mathbf{X}^N|B^{(1)}=b}-\prod P_{\mathbf{X}}||_{TV}
\end{align}

Note that given any fixed $b\in C_B$ the number of Gaussian codewords is $|C_Y|=2^{NR_{Y^{(1)}}}$. 
Also, one can check by the signal model defined in \eqref{eq:regression_star} that the statistical properties of the output vector $\mathbf{X}$ given any fixed sign value $b\in C_B$ does not change. 
Hence, for sufficiently large rates, i.e., $R_{Y^{(1)}}\geq I[\mathbf{X};Y^{(1)}]$, and by soft covering lemma, the term in the summation in \eqref{eq:nearly} vanishes as $N$ grows.
So overall the term shown in \eqref{eq:nearly} vanishes.
This shows that in fact $\gamma_{\mathbf{X}^N}$ \textit{nearly} satisfies the constraints $2)$ and $3)$.
Hence, let's construct another distribution using $\gamma_{\mathbf{X}^N,Y^{(1)},B^{(1)}}$. Define,
\begin{align*}
q_{\mathbf{X}^N,Y^{(1)},B^{(1)}}=\dfrac{1}{|C_B|}(\prod P_{\mathbf{X}})\gamma_{Y^{(1)}|\mathbf{X}^N,B^{(1)}}
\end{align*}

It is not hard to see that such density satisfies $1)-5)$. 
We only need to show that it satisfies $6)$ as well. We have,
\begin{align}
&||q_{\mathbf{X}^N}-\prod P_{\mathbf{X}}||_{TV}\notag\\
&\leq ||q_{\mathbf{X}^N}-\gamma_{\mathbf{X}^N}||_{TV} + ||\gamma_{\mathbf{X}^N}-\prod P_{\mathbf{X}}||_{TV}\notag\\
&\leq ||q_{\mathbf{X}^N,Y^{(1)},B^{(1)}}-\gamma_{\mathbf{X}^N,Y^{(1)},B^{(1)}}||_{TV} + \epsilon_N\label{eq:2nd}\\
&=||q_{\mathbf{X}^N,B^{(1)}}-\gamma_{\mathbf{X}^N,B^{(1)}}||_{TV} + \epsilon_N\label{eq:4th}\\
&=||\dfrac{1}{|C_B|}(\prod P_{\mathbf{X}})-\gamma_{\mathbf{X}^N,B^{(1)}}||_{TV} +  \epsilon_N \label{eq:condition6}
\end{align}

\noindent where $\epsilon_N=||\gamma_{\mathbf{X}^N}-\prod P_{\mathbf{X}}||_{TV}$.
Both terms in \eqref{eq:condition6} vanish as $N$ grows, due to \eqref{eq:nearly} and \eqref{eq:6}, respectively. Note that, \eqref{eq:2nd} is due to \cite[Lemma V.I]{cuff}. Also, \eqref{eq:4th} is due to \cite[Lemma V.II]{cuff}, by considering the terms $p_{\mathbf{X}^N,Y^{(1)},B^{(1)}}$ and $\gamma_{\mathbf{X}^N,Y^{(1)},B^{(1)}}$ as the outputs of a unique channel specified by $\gamma_{Y^{(1)}|\mathbf{X}^N,B^{(1)}}$, with inputs $p_{\mathbf{X}^N,B^{(1)}}$ and $\gamma_{\mathbf{X}^N,B^{(1)}}$, respectively.
\begin{rmk}
\it{
For cases other than the broadcast channel, we know by previous discussions that the input vector $\mathbf{Y}$ follows a mixture Gaussian model, hence, the joint density $p_{\mathbf{X,Y}}$ becomes a mixture Gaussian as well.
However, similar arguments as in the broadcast channel hold to show the achievable region for such cases.
Only this time the lower bounds on the rates is computed under the mixture Gaussian assumption.
}
\end{rmk}

\section{Proof of Lemma \ref{lem:optimal_dumbbell}} \label{app:optimal_dumbbell}
For the channel shown in Figure \ref{fig:B1}, we may write,

\begin{align}
I(&\mathbf{X};\mathbf{B}|\mathbf{Y})=h(\mathbf{X}|\mathbf{Y})-h(\mathbf{X}|\mathbf{Y,B})\notag\\
&=h(\mathbf{X}|\mathbf{Y}) - h(X_1,X_2|Y_1^{(1)},B_1^{(1)})\notag\\
&-h(X_3,X_4|Y_2^{(1)},B_2^{(1)})\label{eq:cond_indep}\\ 
&=h(X_1,X_2|Y_1^{(1)},Y_2^{(1)})+h(X_3,X_4|Y_1^{(1)},Y_2^{(1)})\notag\\
&- h(X_1,X_2|Y_1^{(1)},B_1^{(1)})-h(X_3,X_4|Y_2^{(1)},B_2^{(1)})\label{eq:chain_rule}\\ 
&=h(X_1,X_2|Y_1^{(1)}Y_2^{(1)})+h(X_3,X_4|Y_1^{(1)}Y_2^{(1)})\notag\\ \label{eq:markov_layer}
&- h(X_1,X_2|Y_1^{(1)},B_1^{(1)})-h(X_3,X_4|Y_2^{(1)},B_2^{(1)})\\ 
&=I(X_1,X_2;B_1^{(1)}|Y_1^{(1)}Y_2^{(1)})\notag\\
&+I(X_3,X_4;B_2^{(1)}|Y_1^{(1)}Y_2^{(1)}) \label{eq:split}
\end{align}
\noindent where \eqref{eq:cond_indep} is due to the stated constraints in \eqref{eq:common_info}. Equation \eqref{eq:chain_rule} results using the chain rule and the fact that $(X_1,X_2)-Y_1^{(1)}-Y_2^{(1)}-(X_3,X_4)$ forms a Markov Chain, hence $(X_1,X_2)$ is conditionally independent of $(X_3,X_4)$ given $(Y_1^{(1)},Y_2^{(1)})$.
Note that due to dependency of $B_1^{(1)}$ and $B_2^{(1)}$, we know $h(X_1,X_2|Y_1^{(1)}Y_2^{(1)})\neq h(X_1,X_2|Y_1^{(1)})$, since the latter ignores such dependency.
Hence, to find the optimal solution, one may maximize both terms in \eqref{eq:split} simultaneously, which by definition is equivalent to maximizing the sum $h(X_1,X_2|Y_1^{(1)}Y_2^{(1)})+h(X_3,X_4|Y_1^{(1)}Y_2^{(1)})$. Considering the first term, we know $h(X_1,X_2|Y_1^{(1)}Y_2^{(1)})=\sum_{y_1^{(1)}y_2^{(1)}} p_{Y_1^{(1)}Y_2^{(1)}}(y_1^{(1)}y_2^{(1)}) h(X_1,X_2|y_1^{(1)}y_2^{(1)})$, and due to \cite{capacity} we know the maximum of $h(X_1,X_2|y_1^{(1)}y_2^{(1)})$ happens for uniform conditional density $p(x_1,x_2|y_1^{(1)}y_2^{(1)})$. However, such conditional PDF can be written as conditional PMF $p_{B_1^{(1)}}(b_1^{(1)}|b_1^{(1)}.b_2^{(1)})$, which by previous arguments we may conclude that the optimal solution happens for uniform PMF, which one may easily check that such uniform PMF will be deduced for $(\pi_1,\pi_2)=(1/2,1/2)$. This completes the proof.

\end{document}